\documentclass[]{elsarticle}

\usepackage{lineno,hyperref}
\modulolinenumbers[5]

\usepackage{amsmath,amssymb}
\usepackage{graphicx}
\usepackage{latexsym}
\usepackage{url}
\usepackage{graphics}
\usepackage{color}
\usepackage[normalem]{ulem}
\usepackage{textcomp}
\usepackage{wasysym}
\usepackage{multirow}
\newtheorem{defn}{Definition}

\newcommand{\rt}[1]{{\it #1}}
\newcommand{\U}{\mathcal{U}}
\newcommand{\usr}[1]{u_{#1}}
\newcommand{\UE}{\mathcal{UE}}
\newcommand{\UR}{\mathcal{UR}}
\newcommand{\urt}[1]{\alpha_{#1}}
\newcommand{\UG}{\mathcal{UG}}

\newcommand{\PI}{\mathcal{P}}
\newcommand{\p}[1]{f_{#1}}
\newcommand{\PE}{\mathcal{PE}}
\newcommand{\PR}{\mathcal{PR}}
\newcommand{\prt}[1]{\beta_{#1}}
\newcommand{\PG}{\mathcal{PG}}

\newcommand{\uf}[1]{\phi_{#1}}
\newcommand{\pf}[1]{\psi_{#1}}
\newcommand{\upath}[1]{\langle #1\rangle}
\newcommand{\ppath}[1]{\langle #1\rangle}
\newcommand{\jmpu}[1]{\Circle_{#1}}
\newcommand{\jmpp}[1]{\CIRCLE_{#1}}
\newcommand{\marku}[1]{\triangledown_{#1}}
\newcommand{\markp}[1]{\blacktriangledown_{#1}}
\newcommand{\up}{\rhd}
\newcommand{\pu}{\RHD}

\newcommand{\MU}{\varGamma}
\newcommand{\MP}{\varDelta}
\newcommand{\VU}{V_{U}}
\newcommand{\VP}{V_{P}}
\newcommand{\A}{\rho}
\newcommand{\B}{\varrho}
\newcommand{\g}{\tau}
\newcommand{\model}[1]{\MU,  \MP,  \A,  \B,  #1,  \g}

\newcommand{\nomp}[1]{\mathsf{#1}}
\newcommand{\prop}[1]{\it{#1}}
\newcommand{\own}{\mathsf{own}}
\newcommand{\req}{\mathsf{req}}

\newcommand{\catn}[1]{\lfloor#1\rfloor}
\newcommand{\rh}[1]{\lceil#1\rceil}
\newcommand{\catf}[1]{cf(#1)}

\newcommand{\trva}[1]{t_{#1}}

\journal{Computers \& Security}

\begin{document}

\begin{frontmatter}

\title{A New Access Control Scheme for Facebook-style Social Networks\footnote{
This article is a revised and extended version of~\cite{PZ14} that has appeared in the proceedings 
of the 9th IEEE Conference on Availability, Reliability and Security (ARES 2014). }
}

\author[ad1,ad2]{Jun Pang\corref{cor1}}
\ead{jun.pang@uni.lu}

\author[ad1]{Yang Zhang}
\ead{yang.zhang@uni.lu}

\address[ad1]{University of Luxembourg, Faculty of Sciences, 
Technology and Communication \\
6, rue Coudenhove-Kalergi, L-1359 Luxembourg
}
\address[ad2]{University of Luxembourg,
Interdisciplinary Centre for Security, Reliability and Trust \\
4, rue Alphonse Weicker, L-2721 Luxembourg
}

\cortext[cor1]{Corresponding author.
Tel: +352 466 644 5483,
fax: +352 466 644 35483
}

\addtolength{\abovedisplayskip}{-1ex}
\addtolength{\belowdisplayskip}{-1ex}

\pagestyle{plain}
\begin{abstract}
The popularity of online social networks (OSNs) makes the protection of users' private information
an important but scientifically challenging problem. 
In the literature, relationship-based access control schemes have been proposed to address this problem.
However, with the dynamic developments of OSNs,
we identify new access control requirements which cannot be fully captured by the current schemes.
In this paper, we focus on public information in OSNs and
treat it as a new dimension which users can use to regulate access to their resources. 
We define a new OSN model containing users and their relationships as well as public information.
Based on this model, we introduce a variant of hybrid logic for formulating access control policies.
We exploit a type of category information and relationship hierarchy to further extend our logic for its usage in practice.
In the end, we propose a few solutions to address the problem of information reliability in OSNs,
and formally model collaborative access control in our access control scheme.
\end{abstract}


\end{frontmatter}

\section{Introduction}
\label{sec:intro}

Online social networks (OSNs) are among the most popular web services during the past ten years
and have attracted a huge amount of users all over the world.
For example, Facebook, the leading OSN service, has more than one billion active users monthly.\footnote{\url{http://newsroom.fb.com/}}
OSNs are playing an important role in our daily life by providing a platform for users to present themselves,
articulate their social circles, interact with each other etc.

With the large amount of data maintained in OSN websites,
privacy concerning users' personal information inevitably becomes
an important but scientifically challenging problem.
Access control schemes 
(e.g., see~\cite{S93,SCFY96,A03,AF03,LMW05,BBL05,RCHC09,LLWC12})
are naturally introduced to protect users' private information or resources in OSNs.
They can be used to guarantee that resources are only accessible by the intended users,
but not by other (possibly malicious) users.
Users can control the access to their own information or resources with
access control schemes supplied by OSNs.
The existing schemes, including the ones proposed by the research community,
are mainly \emph{relationship-based}, 
i.e., whether a user is able to access the information depends on the relationship
between him and the owner, e.g., `friends' or `friends of friends'.

Due to their own nature and the development of information and communications technology,
OSNs admit quick and dynamic evolutions.
Many new services and methods for user interaction have emerged.
For instance, users can play online games with friends or find people who share similar interests.
More recently, with the increased popularity of GPS-enabled mobile devices,
OSNs have evolved into geo-social networks -- users can tag posts and
photos with their geographical locations, find nearby friends and post
check-in of some places to share their comments.
OSNs are also emerging as important social media -- 
people use OSNs to publish news, organize events or even seek for emergent help.
For example, Facebook and Twitter play an extremely important role 
during the rescue process for the ``April 2011 Fukushima earthquake'';
and in summer 2014, the ``Ice Bucket Challenge'' 
have achieved a huge success through social media.\footnote{\url{http://en.wikipedia.org/wiki/Ice_Bucket_Challenge}}
(In Section~\ref{sec:motiv}, we will take Facebook as a~typical example
and discuss its developments in the past few years.)

With these evolutions, more information and activities of users
are made available in OSNs.
As a result, new access control schemes are needed to capture these new developments.
Let us illustrate this need by a few scenarios in OSNs.

\begin{itemize}
 \item Someone broke the window of Alice's expensive car and took her purse
       when she parked the car in the area of Montparnasse in Paris.
       Alice publishes a status in the OSN to see if anyone can provide her some clue to find the purse back.
       She doesn't want everyone to know that she has an expensive car, 
       and people who live in other areas or cities won't be able to give her any useful information.
       Therefore, she intends to choose people who live in the Montparnasse area as audiences of her status.
 \item Bob wants to organize a fundraising party for children's rare diseases.
       He doesn't want to make this event public as certain sensitive information of the participants can be leaked, 
       e.g., it is possible that some participants' family members may suffer from the disease.
       Instead, Bob only wants people who are linked with a certain number of charities 
       (through donations, volunteering, etc) as him to attend the party.
 \item Charlie has some friends who work at the rival company of his own employer.
       These friends invited him to attend the party organized by their company.
       Charlie publishes a photo taken at the party.
       Apparently, it is not a good idea for his colleagues and boss to see this photo.
       Thus Charlie wants no one but his friends who work at this rival company to see it.
\end{itemize}
In relationship-based schemes, a resource owner cannot exploit any other
information but user relationships between him and the requester when defining access control policies.
Therefore, the above requirements cannot be fully and precisely
formulated in the current schemes proposed in the literature.

\smallskip
\noindent\textbf{Contributions and Outline.}
In order to solve the identified problems,
we propose a new access control scheme for OSNs.
We focus on public information existing, e.g., in Facebook (Section~\ref{sec:motiv}),
and show that it can be used to group users based on their attributes, common interests and activities.
Public information can thus be considered as a~new dimension for users to regulate access to their resources.
As a consequence, we propose a new OSN model containing
both a user graph and a public information graph (Section~\ref{sec:model}).
We then extend a hybrid logic~\cite{BFSH12} to express this type of access control policies (Section~\ref{sec:logic}).
The expressiveness of our scheme is extensively discussed
through a number of real-life scenarios (Section~\ref{sec:example}).
We further identify two special semantic relations, i.e., \emph{category relation} among public information and \emph{relationship hierarchy},
which allow us to express certain types of policies in a concise way (Section~\ref{sec:category} and Section~\ref{sec:hierarchy}).
To address the problem of information reliability in OSNs,
we propose to add endorsement and trust into our policy formulas (Section~\ref{sec:trust}).
In addition, we formally model the collaborative access control in Section~\ref{sec:collaborative}
within our new access control scheme.

After the introduction, we give a brief overview of related work in Section~\ref{sec:related}.
Section~\ref{sec:compare} compares our access control scheme
with existing schemes in the literature.
We discuss several issues related to our scheme in Section~\ref{sec:discuss}
and conclude our paper with some future work in Section~\ref{sec:conclusion}.

\section{Related Work}
\label{sec:related}

Relationship-based access control, driven by OSNs, was first advocated in~\cite{G07} 
and defined as an access control paradigm based on interpersonal relationships.
Carminati et al.\ proposed the first relationship-based access control model in~\cite{CFP09},
where the relationships between the qualified requester and the owner are interpreted into three aspects, 
i.e., relationship type, depth and trust level. 
In~\cite{CFHKT09}, the authors used semantic web technology including OWL and SWRL
to extend the model of~\cite{CFP09}. 
They also proposed administrative and filtering policies
which can be used for collaborative and supervising access control, respectively.
Fong et al.\ proposed an access control scheme for Facebook-style social networks~\cite{FAZ09},
in which they model the access control procedure as two stages. 
In the first stage, the requester has to find the owner of the target resource;
then in the second stage, the owner decides whether the authorization is granted or not. 
Their access control policies are mainly based on the relationships between the requester and the owner. 
Moreover, they proposed several meaningful access control policies
based on the graph structure of OSNs, such as $n$-common friends and clique. 
In~\cite{F11b}, Fong introduced a modal logic to define access control policies for OSNs. 
Later Fong and Siahaan~\cite{FS11} improved the previously proposed logic
to further support policies like $n$-common friends and clique. 
In~\cite{BFSH12}, the authors adopted a hybrid logic to describe policies
which eliminates an exponential penalty in expressing complex relationships such as $n$-common friends.
This hybrid logic is expressive and has been adopted by several other works~\cite{TF14,TFM14,CPZ15} for specifying access control policies.
A visualization tool for evaluating the effect of access control configurations is designed in~\cite{AF12},
with which a user can check which other users within a certain distance to him can view his resources.
Cheng et al.\ proposed a rich OSN model in~\cite{CPS12}. 
In their work, not only users but also resources are treated as entities
and actions performed by users are considered as relationships in OSNs.
As more information are incorporated in their model, 
many new access control policies can be expressed (more details can be found in Section~\ref{sec:compare}).
Their model supports administrative and filtering policies as proposed in~\cite{CFHKT09}.
Recently, Crampton and Sellwood~\cite{CS14} generalized relationship-based access control to other systems than social networks,
they proposed path logic conditions for specifying policies and adopt principle matching for policy evaluation.
Besides models, several security protocols based on cryptographic techniques are proposed 
to enforce relationship-based access control policies, e.g., see~\cite{CF08,DVSG08,CF09,FS09,MPGP09,XCF11,BMP11,NCDAS13,PZ15b}.

As a shared platform, resources in OSNs may be co-owned by a number of users.
Thus, collaborative access control also plays an essential role in protecting privacy.
A game theoretical method based on the Clarke-Tax mechanism 
for collective privacy management was proposed by Siquicciarini et al.~\cite{SSP09}.
Sun et al. proposed a different approach by combining trust relations in OSNs 
and preferential voting schemes~\cite{SZPAM11,SZPAM12}. 
Ahn et al. introduced a multiparty access control model in~\cite{HAJ13}. 
In addition, they developed a policy specification scheme and
a voting based conflict resolution mechanism.
Photo tagging is the~most common service relevant to collaborative access control.
The authors of~\cite{BL10,KLMSUB12} have investigated users' privacy concerns about this service and
proposed principles for designing better collaborative access control schemes.
Besides interaction, users' private information can be leaked through third party applications.
A privacy-by-proxy design for social network APIs was developed by Felt and Evans~\cite{FE08}. 
Singh et al.~\cite{SBL09} proposed a privacy-preserved application platform, i.e., \verb'xBook', 
which integrates information flow model to control what applications can do with users' information.
An access control scheme for third party applications was developed~\cite{SSAK12},
where applications are required to adapt users' specifications on their own data.

\section{Motivation}
\label{sec:motiv}

An OSN provides users with some typical services,
such as users can build their profiles and establish social relationships with each other.
Moreover, an OSN also provides a platform for users to socialize and interact with each other.
In the following, we first give a brief overview of the developments of Facebook --
one of the most popular OSN services in the world.
After that, we discuss public information in Facebook and its potential usage in access control.

\subsection{Facebook}
In Facebook, each user is affiliated with a personal profile
that contains his basic information (e.g., age, gender and nationality),
work and education background, living places and so on.
A user's hobbies (e.g., sports, movies and music) are articulated in `Likes'; 
places he has been to are marked in `Map'.
Besides personal representation,
Facebook also allows a user to establish friend relations with others.
In addition, Facebook recommends friends for users based on common friends, same hometown or similar interests.
A user can organize his friends into different groups, or named friend list.
Moreover, Facebook also automatically create lists (smart list) for users based on their work, living area, school and family.

Facebook is not only a website storing users' personal information and social relations,
but also a platform for users to interact with each other.
A~user can directly communicate with his friends by sending messages; 
he can tag his friends in photos and posts.
Two friends can interact through Facebook applications such as games.
All activities performed by a user are organized chronologically in his `Timeline'
through which other users as well as the user himself can check his past activities conveniently.
A user receives his friends' news on `Newsfeed'.
When he finds something interesting, he can further perform actions,
such as `like', `share' and `comment', on it.
Users on Facebook can also establish public groups
and organize events such as birthday parties, meetings and conferences,
and invite other users to attend, like or share these groups or events.

In January 2013, Facebook publishes a new product called Graph Search, 
a~search engine based on users' data.\footnote{\url{https://www.facebook.com/about/graphsearch}}
It allows users to explore more information about daily life,
find people who share common interests or live in the same city,
discover new restaurants and music, and so on.
Through Graph Search, a user can directly acquire information from his friends' data 
without visiting their personal pages.
For example, if a user types in ``photos by my friends'', 
he will get a page containing all photos uploaded by his friends.
Since Graph Search is a personalized search engine,
for the same query different users will get different results.

When a user wants to publish or share a resource (a photo or a post),
Facebook provides him an audience selector to let him decide who can view this resource.
This audience selector is the access control implementation of Facebook 
and it supports five different modes
including `public', `friends', `friends except acquaintances', `only me' and `custom'. 
In the last mode, a user can choose the eligible requester to be (or not to be)
a single user or a specific group (through friend list).
The selector also supports smart lists in Facebook,
which group a user's friends according to `work', `school', `family' and `city'
using the information added to the `Education' and `Work' and `Current City' sections of their profile. 

\subsection{Public Information and Access Control}
Besides users' information,
Facebook imports knowledge of external sources, 
e.g., Wikipedia and Bing map, into its system to formalize another type of entities.
We name them \emph{public information}.
A lot of entities in the real world are modeled as public information,
e.g., countries, history events or public figures.
Public information are mainly used as common reference points of users' information,
through which a user can find other users in Facebook
with similar background, hobbies, experiences, etc.
For example, a~user can find his schoolmates
through the public information of the college that he has attended.

Each public information is affiliated with 
a content that is normally extracted from external sources.
Similar to users, public information are also connected with each other
and links among them are based on their contents.
For example, if Wikipedia articles of two charities are connected,
then their public information in Facebook are connected as well.
Besides, there exist many different links between users and public information. 
Some of these connections are based on user profiles, 
e.g., if a~user specifies his employer in his profile, 
then he is linked with this employer's public information.
Others are computed by Facebook through mining users' data.
For example, if a user posts a status labeled with a location,
then the user is connected with the location's public information.

In addition to facilitate users' interaction,
it is possible to use public information in expressing access control requirements.
For example, in the first scenario as discussed in Section~\ref{sec:intro}, 
the requester has to be linked to the location where the car was parked; 
in the second one, the requester needs to be linked with the owner through some charity organizations;
in the third one, the requester is asked to be connected with the owner through 
not only a friendship but also their employers' connection.
Here, the location, charities as well as companies can all be modeled as public information in OSNs.

\begin{figure*}[!t]
\centering
\includegraphics[scale=0.5]{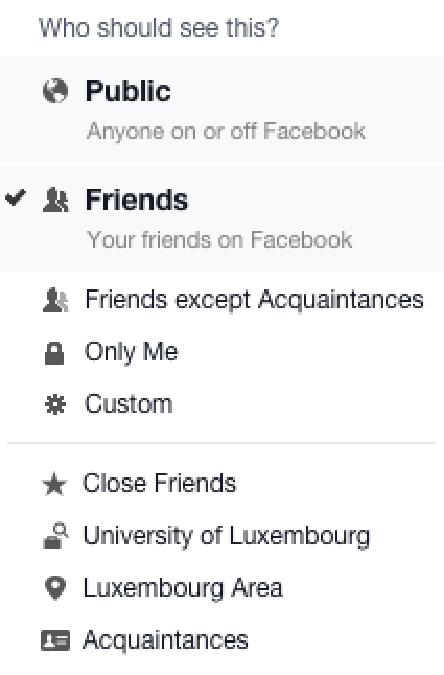}\hspace{4mm}
\includegraphics[scale=0.6]{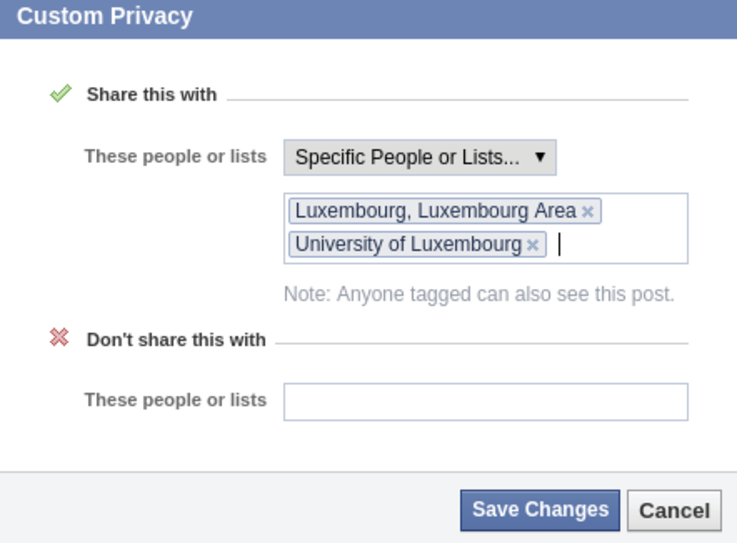}
\caption{Access control with smart list in Facebook.\label{fig:smartlistaccess}}
\end{figure*}

All the above access control requirements are meaningful
and in line with the recent developments of OSNs.
However, the current access control schemes proposed in the literature mainly focus on relationships among users,
public information are not taken into account.
On the other hand, Facebook already allows users to define policies with some simple public information.
As shown in Figure~\ref{fig:smartlistaccess}, 
a user can define a policy to allow users who lives in the same area
or work at the same university as him to view his photo through smart lists.
However, this function is still ad hoc, scenarios proposed in Section~\ref{sec:intro} cannot be fully captured.
Therefore, in this paper we propose a new access control scheme,
in which policies can be expressed based on both users and public information, and their relationships.

\section{A Model of Online Social Networks}
\label{sec:model}

Our OSN model contains information of 
(1) users and their social relationships, 
(2) public information and their connections, and 
(3) links between users and public information. 
Public information and users are essentially two different concepts --
public information are imported from external databases (in most cases), 
and they cannot perform actions and establish relationships with each other as users;
relationships among public information are also extracted from external sources.
Therefore, we treat public information and users separately. 
We model an OSN as a tuple $(\UG, \PG, \A, \B)$. 
A user graph is denoted by $\UG$, and it depicts users and their relationships. 
A public information graph is denoted by $\PG$,
which represents all public information and connections among them.
Two maps, i.e., $\A$ and $\B$, store links between users and public information.

\subsection{User Graph}
The set $\U$ contains all users in an OSN.
Each user is affiliated with some basic information which are treated as attributes of the user.
We use $\UR=\{\urt{1},\urt{2},\ldots,\urt{k}\}$ to denote a (finite) set of relationship types supported in the OSN.
The semantics of each relationship type is defined as $\urt{i}\subseteq \U \times \U$.
If user $\usr{a}$ is in a $\urt{i}$ relationship with user $\usr{b}$, then we write $(\usr{a},\usr{b})\in \urt{i}$.
For each relationship type $\urt{i}\in\UR$, there exists its reverse relationship type, 
e.g., if $\urt{i}$ stands for $\rt{husbandof}$, then its reverse is $\rt{wifeof}$. 
We use $\urt{i}^{\!-\!1}\in \UR$ to denote the reverse of $\urt{i}$. 
Moreover, if $\urt{i}=\urt{i}^{\!-\!1}$, then $\urt{i}$ is a \emph{symmetric relationship},
e.g., $\rt{friend}$ is a typical symmetric relationship.
User graph $\UG$ is a directed graph denoted as $(\U, \UE)$,
where every user in the OSN is a node and the set of edges, i.e., $\UE$, is defined as
$\{(\usr{a},\usr{b},\urt{i})\mid \usr{a}, \usr{b}\in \U\mbox{ and }(\usr{a},\usr{b})\in \urt{i}\}$. 

\subsection{Public Information Graph}
As we introduced in Section~\ref{sec:motiv},
public information are also linked as together, such as Paris is linked with France.
Therefore, we model public information as a graph.
We use the set $\PI$ to denote all public information that are extracted 
from external databases, such as Wikipedia and some geography databases (such as Bing).
Each public information $\p{c}$ has its own attributes.
We use $\PR=\{\prt{1},\prt{2}, \ldots,\prt{\ell}\}$ to denote a (finite) set of 
relationship types on public information.
Each relationship type $\prt{j}$ can be semantically defined as $\prt{j}\subseteq \PI\times \PI$.
If $\prt{j}$'s reverse relationship type exists, it is denoted by $\prt{j}^{-1}$.
Public information graph is formally denoted as $\PG=(\PI, \PE)$, 
where $\PI$ is the set of nodes and $\PE$ is defined as 
$\{(\p{c}, \p{d}, \prt{j})\mid \p{c}, \p{d}\in \PI \mbox{ and } (\p{c},\p{d})\in \prt{j}\}$.
%

\subsection{Links between $\UG$ and $\PG$}
There are a lot of links between users and public information.
For example, a user is linked with the language he speaks and the city he lives in.
As the OSN is modeled as $\UG$ and $\PG$, we define two maps, 
i.e., $\A$ and $\B$, between them to describe their connections:
\[
\begin{array}{cccc cccc}
\A: & \U &\to & 2^{\PI} & \hspace{2mm}\mbox{and}\hspace{3mm}
\B: & \PI &\to &2^{\U}.
\end{array}
\]
For a user $\usr{a}\in \U$, $\A(\usr{a})$ is a subset of the nodes in $\PG$ that are related to $\usr{a}$. 
The map $\A(\usr{a})$ may contain a lot of different types of public information, 
such as museums, universities, pop stars, etc, 
which are computed by the OSN with the information that $\usr{a}$ provides. 
For a public information $\p{c}\in \PI$, $\B(\p{c})$ gives all the users in $\UG$ 
who have been involved in activities or have information related to $\p{c}$. 
How to compute $\A$ and $\B$ is not the focus of this paper,
we assume that $\A$ and $\B$ always give us the right results.
In practice, it is desirable to have more fine-grained links between users and public information.
With respect to this, the two maps $\A$ and $\B$ can be further refined
to reflect how precisely a user and a piece of public information is connected.

\subsection{An Example}
A sample OSN model is shown in Figure~\ref{fig:model},
whose left side is a $\UG$ and right side is a $\PG$.
Edges in the graph with double arrows imply that the relationships are symmetric.\footnote{
For the sake of simplicity, we omit some edges in the figure, e.g., the edge from Danny and Eve to represent
the relationship `husbandof'.}
For example, Alice and Bob are friends; Company~A and Company B are rivals.
The dash lines between users and public information reflect the links between $\UG$ and $\PG$,
which are formally captured by the two maps $\A$ and $\B$.
(The part contained in the dashed box in the right-bottom corner will be discussed in Section~\ref{sec:category}.)

\section{A Hybrid Logic}
\label{sec:logic}

In~\cite{BFSH12}, a hybrid logic is used to define access control policies for OSNs. 
We adopt their logic and additionally introduce a~new type of formulas $\pf{}$.
With such formulas, we can define policies based on information in $\PG$. 
Moreover, two new logic operators, i.e., $\up$ and $\pu$, are introduced to connect formulas on $\UG$ and $\PG$, respectively. 
In this way, we can combine resources and their relations from both $\UG$ and $\PG$
to specify new and expressive access control policies (see examples in Section~\ref{sec:example}). 

\subsection{Syntax}
The syntax of our hybrid logic is given below,
and its semantics will be discussed in the next section.
\[
\arraycolsep=2pt
 \begin{array}{ccccccccccccccccc}
    s &::= & m & \mid & x \\
    t &::= & n & \mid & y \\
  \uf{} &::= & s & \mid & p & \mid & \neg \uf{} & \mid & \uf{1}\wedge \uf{2} & \mid & \upath{\urt{i}}\uf{} & \mid & \jmpu{s}\uf{} &
  \mid & \marku{x}\uf{} & \mid & \up\pf{}   \\
  \pf{} &::= & t & \mid & q & \mid & \neg \pf{} & \mid & \pf{1}\wedge \pf{2} & \mid & \ppath{\prt{j}}\pf{} & \mid & \jmpp{t}\pf{} & 
  \mid & \markp{y}\pf{} & \mid & \pu\uf{}\\
\end{array}
\]
In our logic, there are mainly two types of formulas:
the user formulas $\uf{}$ manipulate information on the user graph $\UG$,
while the public information formulas $\pf{}$ are defined on $\PG$.
Three kinds of atoms are supported in our logic, 
i.e., nominals ($m$ and $n$), variables ($x$ and $y$) and proposition symbols ($p$ and $q$). 
Nominal $m$ represents the name of a user in $\UG$, e.g., Alice,
while $n$ represents the name of a public information in $\PG$, e.g., Paris.
Propositional symbol $p$ is used for specifying the attributes of users in $\U$
and similarly $q$ is used for public information in $\PI$. 
For example, $p$ (i.e., $\prop{IsMale}$) can specify users who are male
and $q$ (i.e., $\prop{IsCity}$) can specify those publication information representing a city. 
Atoms $m$, $x$ and $p$ are used in user formulas $\uf{}$,
while $n$, $y$ and $q$ are used in public information formulas $\pf{}$. 
Negation $\neg$ and conjunction $\wedge$ have their usual meanings and
can be used to define disjunction $\vee$.
Therefore, we also use $\vee$ in both $\uf{}$ and $\pf{}$.
$\upath{\urt{i}}$ and $\ppath{\prt{j}}$ are two modal logic operators.
As described in Section~\ref{sec:model}, 
symbols $\urt{i}$ and $\prt{j}$ represent the relationship types in $\UG$ and $\PG$, respectively. 
Hybrid logic operator $\jmpu{}$ can be used either with a nominal or variable, 
while $\marku{}$ can only operate on variables. 
The same holds for $\jmpp{}$ and $\markp{}$. 
Two new logic operators, i.e., $\up$ and $\pu$, 
are used to connect the two types of formulas $\uf{}$ and $\pf{}$ together.
They allow the specification of access control policies
based on both information from the user graph and the public information graph.

\subsection{Semantics}
Our model for evaluating access control policy formulas contains six parts, 
i.e., $\model{{\it cur\_n}}$, where $\MU=(\UG, \VU)$ and $\MP=(\PG, \VP)$.
$\VU$ is a map between atoms (either $m$ or $p$) and users in $\UG$,
$\VU(m)$ is a set that contains only one user in $\UG$ whose name is $m$
and $\VU(p)$ is a set of users that have the attribute as specified by $p$.
For example, $\VU({\it Alice})$ refers to a singleton containing the node of Alice in $\UG$.
Similarly, we can define $\VP(n)$ and $\VP(q)$. 
As introduced in Section~\ref{sec:model}, $\A$ and $\B$ maintained by the OSN connect users and public information. 
Node $\it cur\_n$ refers to either a user $\usr{a}$ in $\UG$ or a public information $\p{c}$ in $\PG$. 
Valuation $\g$ stores all the maps from variables in the policy formula to vertices in either $\UG$ or $\PG$. 
When there is a new map from $x$ to $\usr{a}$ ($y$ to $\p{c}$) added to $\g$, 
we write $\g[x\mapsto \usr{a}]$ ($\g[y\mapsto \p{c}]$).

We use satisfaction relation $\model{\usr{a}}\vDash \uf{}$ to describe the meaning of user formula~$\uf{}$.
\[
\arraycolsep=2pt
 \begin{array}{lll}
  \model{\usr{a}} \vDash x& \mbox{iff}& \usr{a}=\g(x)\\
  \model{\usr{a}} \vDash m& \mbox{iff}& \VU(m)=\{\usr{a}\}\\
  \model{\usr{a}} \vDash p& \mbox{iff}& \usr{a}\in \VU(p)\\
  \model{\usr{a}} \vDash \neg\uf{}& \mbox{iff}& \model{\usr{a}}\nvDash \uf{}\\
  \model{\usr{a}} \vDash \uf{1}\wedge\uf{2} & \mbox{iff}& \model{\usr{a}}\vDash \uf{1}\wedge\model{\usr{a}}\vDash \uf{2}\\
  \model{\usr{a}} \vDash \upath{\urt{i}}\uf{}& \mbox{iff}& \exists \usr{b}\in \U \mbox{ s.t. } (\usr{a},\usr{b})\in \urt{i}
\wedge\model{\usr{b}} \vDash \uf{}\\
  \model{\usr{a}} \vDash \jmpu{m} \uf{} & \mbox{iff}& \model{\usr{b}} \vDash \uf{} \mbox{ where } \VU(m)=\{\usr{b}\}\\   
  \model{\usr{a}} \vDash \jmpu{x} \uf{} & \mbox{iff}& \model{\g(x)} \vDash \uf{}\\
  \model{\usr{a}} \vDash \marku{x}\uf{} & \mbox{iff}& \MU,\MP,\rho,\varrho,\usr{a},\g[x\mapsto \usr{a}] \vDash \uf{}\\      
  \model{\usr{a}} \vDash \up \pf{} &      \mbox{iff}& \exists \p{c}\in \A(\usr{a})\mbox{ s.t. } \model{\p{c}}\vDash \pf{}
 \end{array}
\]
The first three relations express the meaning of atoms. 
When $\uf{}$ is a~single variable $x$, it holds if and only if when $\g$ contains a map from $x$ to $\usr{a}$. 
If $\uf{}$ is a~single nominal or propositional symbol,
it is true if and only if when $\usr{a}$ is in the set defined by $\VU$. 
When several modal logic operators ($\upath{\urt{i}}$) are aligned sequentially in a formula,
they can represent a~\emph{relationship path}, 
e.g., user can define a~policy to regulate that only `friends of friends' can access his resource.

The hybrid logic operator $\jmpu{s}\uf{}$ jumps to the node that $s$ refers to in $\UG$,
and $\marku{x}\uf{}$ adds a~map from $x$ to $\usr{a}$ into $\g$.
The new operator, i.e., $\up \pf{}$, links a user formula $\uf{}$ with a public information formula $\pf{}$ -- 
it maps the current node $\usr{a}$ in $\UG$ to a set of public information in $\PG$ that are related to this user. 
If there is one public information in $\A(\usr{a})$ satisfying $\pf{}$, then the formula $\up \pf{}$ holds.

In the following, we give the meaning of public information formulas $\pf{}$. 
\[
\arraycolsep=2pt
\begin{array}{lll}
  \model{\p{c}} \vDash y& \mbox{iff}& \p{c}=\g(y)\\
  \model{\p{c}} \vDash n& \mbox{iff}& \VP(n)=\{\p{c}\}\\
  \model{\p{c}} \vDash q& \mbox{iff}& \p{c}\in \VP(q)\\
  \model{\p{c}} \vDash \neg\pf{}& \mbox{iff}& \model{\p{c}} \nvDash \pf{}\\
  \model{\p{c}} \vDash \pf{1}\wedge\pf{2} & \mbox{iff}& \model{\p{c}} \vDash \pf{1} \wedge \model{\p{c}}\vDash \pf{2}\\
  \model{\p{c}} \vDash \ppath{\prt{j}}\pf{}&\mbox{iff}& \exists \p{d}\in \PI \mbox{s.t.} (\p{c},\p{d})\in \prt{j} \wedge \model{\p{d}}\vDash \pf{}\\
  \model{\p{c}} \vDash \jmpp{n} \pf{} &\mbox{iff}& \model{\p{d}}\vDash \pf{} \mbox{ where } \VP(n)=\{ \p{d}\}\\
  \model{\p{c}} \vDash \jmpp{y} \pf{} &\mbox{iff}& \model{\g(y)}\vDash \pf{}\\
  \model{\p{c}} \vDash \markp{y}\pf{} &\mbox{iff}& \MU,\MP,\rho,\varrho,\p{c},\g[y\mapsto \p{c}]\vDash\pf{}\\
  \model{\p{c}} \vDash \pu \uf{} &\mbox{iff}& \exists \usr{a}\in \B(\p{c}) \mbox{ s.t.}\model{\usr{a}}\vDash \uf{}
 \end{array}
\]
It is easy to find that the semantics of public information formulas resembles the user formulas.
Therefore, information in $\PG$ can be used in access control policies in a same way as in $\UG$. 
When the evaluation process encounters the operator $\pu\uf{}$,
the public information node $\p{c}$ is mapped to users that are related to it in $\UG$. 
If $\uf{}$ holds at one of these users, then the formula $\pu\uf{}$ is true.

Note that, by combing the user formula $\up \pf{}$ with propositions,
we can link a user to a more specific set of public information.
We write $\up_{q}\pf{}$ for $\up(q \wedge \pf{})$ and its meaning can be reinterpreted as:
\[
\arraycolsep=2pt
 \begin{array}{lll}
 \model{\usr{a}} \vDash \up_q \pf{}& \mbox{iff}& \exists \p{c}\in \A(\usr{a}) \cap \VP(q) \mbox{ s.t. } \model{\p{c}}\vDash \pf{}
 \end{array}
\]
Similarly, we can define $\pu_{p}\uf{}$ as $\pu(p \land \uf{})$ and formulate its semantics.

\subsection{Expressing Access Control Policies}
In general, there are four elements in an access control scenario, 
i.e., a~requester, a~target, an~action and access control policies. 
More precisely, the requester tries to perform an action on the target, 
whether he succeeds or not depends on the access control policies defined for the target. 

\begin{itemize}
 \item \textbf{Owner and requester.}
 Both the owner and requester are users in the social network,
 and we use free variables $\own$ and $\req$ to represent the owner
 of the resource and the~requester in the formula.
 
 \item \textbf{Target.} With multiple services supported by the OSN, a~target can be a user or a resource. 
 For example, a requester can request to chat with a user or view one of his photo.
 For the sake of simplicity, we assume that the target can only be a~resource owned by some user. 
 
 \item \textbf{Access control policies.} Normally, a~user can define an access control policy for the resources that he owns.
 But in some cases, the access of a resource is decided by several users.
 For example, for a photo that is tagged with several users, each of them should have the right to decide who can view this photo.
 This is the subject of collaborative (or multi-party) access control management, e.g.,
 see~\cite{CPS12,SSP09,SZPAM11,SZPAM12,HAJ13}.
 For now, we assume that a~resource is attached with only one access control policy 
 that is defined by its owner.
 In Section~\ref{sec:collaborative}, we will show how to support collaborative access control
 within our access control scheme.
 
 \item \textbf{Action.} As introduced in Section~\ref{sec:motiv}, 
 a user can perform multiple actions in OSNs, such as `view', `comment', `tag' and `share'.
 The only action we consider here is `view' and other actions are affiliated with it, 
 i.e., when a~user is able to view a~resource published by another user, 
 he can comment or share it as well.
\end{itemize}
In OSNs, both the requester and the~owner are users. 
We restrict that an access control formula has to start with either an owner or a requester,
i.e., policy formulas are in the form either $\jmpu{\own}\uf{}$ or $\jmpu{\req}\uf{}$.

\subsection{Model Checking}
Given an OSN model $(\UG, \PG, \A, \B)$ and an access control policy 
expressed in our hybrid logic as a formula $\uf{}$,
the satisfaction of $\model{\usr{a}} \vDash \uf{}$ 
with $\g[\own\mapsto \usr{a}, \req\mapsto \usr{b}]$, $\MU=(\UG, \VU)$ and $\MP=(\PG,\VP)$
is formulated as a local model checking problem by Bruns et al.~\cite{BFSH12}.
Except for the user graph $\UG$,
our OSN model captures public information and their relationships.
Moreover, our logic essentially extends the one of~\cite{BFSH12}
with public information formulas $\pf{}$ defined on $\PG$ and
two new operators $\up$ and $\pu$
connecting user formulas and public information formulas.
In principle, we can reuse the model checking algorithm of Bruns et al.~\cite{BFSH12}.
As formulas of the form $\up\pf{}'$ or $\pu\uf{}'$
explore the links between $\UG$ and $\PG$, we need to treat them differently. 
A formula $\up \pf{}'$ maps the current node ($cur\_n$) in $\UG$ to a
a set of public information in $\PG$. 
As long as there is one public information in $\A(cur\_n)$ satisfying $\pf{}$, then $\uf{}$ holds.
The formula $\pu\uf{}'$ is defined similarly.
To check them, we can develop a sub-routine similar to {\tt MCmay} of Bruns et al.~\cite{BFSH12},
which first computes the set of all public information (users) related to a specific user (public information)
and then iterate through the set until one of them makes the connected formula $\pf{}'$ ($\uf{}'$) hold
on $\PG$ ($\UG$). For formulas $\up (q\land\pf{}')$ and $\pu(p\land\uf{}')$ as discussed in Section~\ref{sec:logic},
we can further reduce the size of the computed set by using propositions $p$ and $q$
to improve the efficiency in model checking.

\section{Example Policies}
\label{sec:example}

\begin{figure*}[t!]
\centering
\hspace{-6mm}
\includegraphics[scale=0.31]{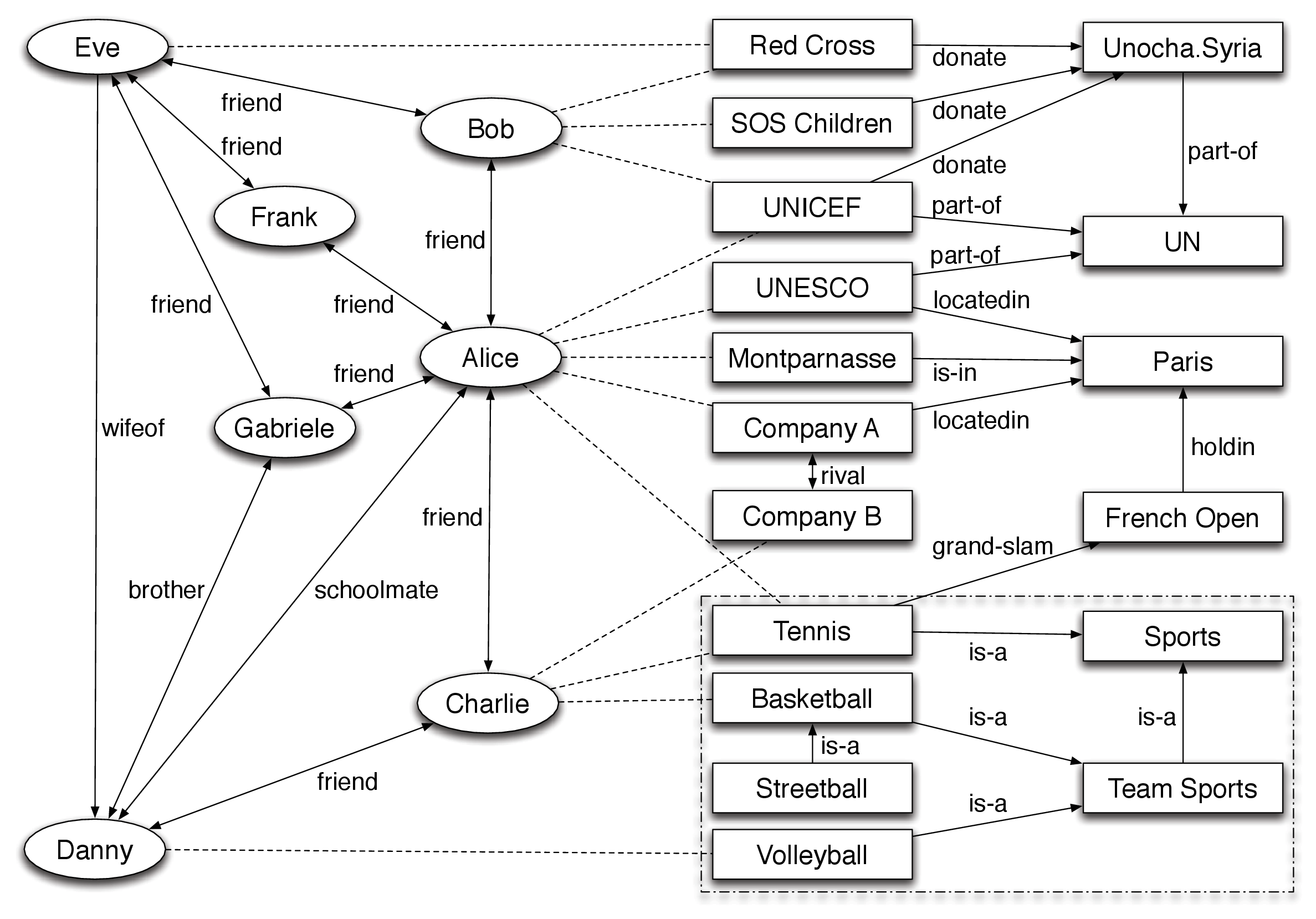}
\caption{A sample OSN model.\label{fig:model}}
\end{figure*}

In order to show the expressiveness of our new scheme based on the OSN model,
we design several real-life scenarios and give their corresponding formulas in our logic. 
We use the OSN model depicted in Figure~\ref{fig:model},
and we assume that valuation $g$ contains two maps $\own\mapsto \usr{o}$ and $\req\mapsto \usr{r}$, 
where $\usr{o}, \usr{r}\in \U$ are the~owner and the~requester, respectively. 

\smallskip
\noindent\textbf{Scenario 0.}
We first show how to express the policy related to user relationships.
Suppose that Eve defines a policy on a certain resource 
to regulate that the qualified requesters can only be her friends or friends of friends.
The policy formula can be written as follows:
\[
\jmpu{\own}(\upath{\rt{friend}} \req\vee \upath{\rt{friend}}\upath{\rt{friend}}\req).
\]
The hybrid logic operator $\jmpu{\own}$ drives the formula to start at Eve.
The requirement ``friends of friends'' is achieved by aligning $\upath{\rt{friend}}$ twice
which forms a~relationship path of length two.
In Figure~\ref{fig:model}, 
Bob, Frank and Gabriele can view the resource because they are friends of Eve, 
Alice is also eligible since she is one of Eve's friends of friends.

To restrict the access to the photo, except for her friends,
Eve regulates that the qualified requester should have at least three common friends with her. 
The policy formula is written as
\[
\jmpu{\own}(\upath{\rt{friend}}\req \vee\upath{\rt{friend}}_3\req).
\]
This is the `$n$-common friends' -- 
one of the topology-based access control policies defined in~\cite{FAZ09} --
$\upath{\rt{friend}}_3$ expresses `at least three different friends' in the formula. 
In~\cite{BFSH12}, the authors show how to implement this policy
with the logic operators $\marku x$ and $\jmpu{s}$, we omit the details here.
In Figure~\ref{fig:model}, as Alice has three common friends with Eve, she can still view the photo.

Next, we illustrate the usage of public information by defining access control policies 
for four different scenarios.
In the first scenario, public information are used to describe an attribute of the qualified requester.
While in the second and third scenarios, the owner and the requester are linked through public information.
In addition, the third scenario needs the owner and the requester to be connected through the user relationship as well.
In the fourth scenario (not discussed in Section~\ref{sec:intro}), the owner and the requester are linked
through a path composed by both users and public information.

\smallskip
\noindent\textbf{Scenario 1.}
Let us recall the first access control scenario discussed in Section~\ref{sec:intro}, 
which exploits the information in $\PG$. 
Alice publishes a status to find a witness who lives in or visited the area where her car was broken into, 
i.e., Montparnasse in Figure~\ref{fig:model}.
The policy is formulated as
\[
 \jmpu{\req}\up\nomp{Montparnasse}.
\]
The operator $\up$ links $\UG$ with $\PG$, 
as introduced in Section~\ref{sec:logic}, we can use $\up_{\prop{IsLocation}}$ to make the map more precisely.
$\nomp{Montparnasse}$ in the formula is a nominal, 
$\VP(\nomp{Montparnasse})$ is the node that represents Montparnasse in $\PG$.
Here, the requester's connection with Montparnasse can be treated as one of his attributes.

In order to get more information, Alice may enlarge the searching area to the whole city, 
i.e., Paris in Figure~\ref{fig:model}.
We assume that a user can only be linked to a place's public information, but not to a city's public information.
For example, a~user's photo can be labeled with any street or square of a city,
but not the city itself. 
The~policy can then be written as 
\[
 \jmpu{\req}\up_{\prop{IsLocation}}\ppath{\rt{is\mbox{-}in}}\nomp{Paris}.
\]
Here, $\ppath{is\mbox{-}in}$ represents a 1-depth relationship path in $\PG$.
Depending on the policy, the length of the path can be arbitrary.
Note that the requester's connection with Paris can be also formalized as an attribute.
However, in this way, each user will be affiliated with a huge number of attributes in the model 
which may not be an ideal solution.

\begin{figure*}[!t]
\centering
\includegraphics[scale=0.28]{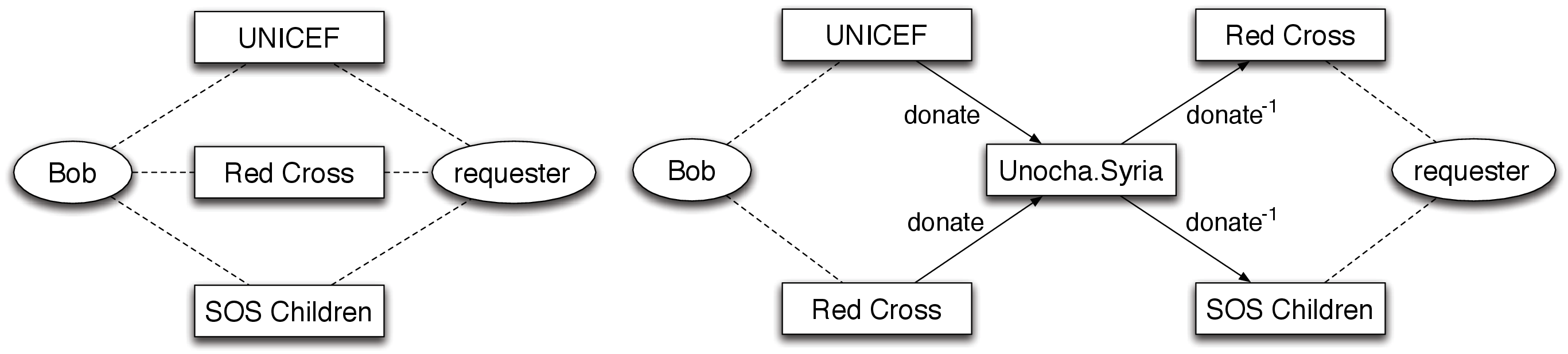}
\caption{Connections between Bob and qualified requesters.\label{fig:cp}}
\end{figure*}

\smallskip
\noindent\textbf{Scenario 2.}
In this scenario (the second one in Section~\ref{sec:intro}), 
Bob wants to use the OSN to organize a fundraising party for children's rare diseases.
He intends to let people who are affiliated with at least a certain number, 
such as three, of different charities as himself to access the event page.
The policy is defined as follows.
\[
 \begin{array}{l}
  \jmpu{\own}\up_{\prop{IsCharity}}\markp{y_1}\pu(\req \wedge \\
  \jmpu{\own}\up_{\prop{IsCharity}}\markp{y_2}(\neg y_1 \wedge \pu(\req \wedge\\
  \jmpu{\own}\up_{\prop{IsCharity}}\markp{y_3}(\neg y_1 \wedge \neg y_2 \wedge \pu \req))))
 \end{array}
\]
The left part of Figure~\ref{fig:cp} depicts an example of three charities (`UNICEF', `Red Cross' and `SOS Children's Villages')
in $\PG$ needed between a qualified requester and Bob.
It can be thought as a public information version of `3-common friends' policy in $\UG$.
Three variables, i.e., $y_1$, $y_2$ and $y_3$, mark three charities that Bob is linked with;
the conjunction of their negative forms, i.e., $\neg y_1$ and $\neg y_1\land \neg y_2$,
in the formula makes sure that these three charities are different.

With our logic, more complicated policies can be achieved based on the information of $\PG$.
Suppose that Bob wants to organize another fundraising party 
for homeless children in Syria during its current civil war.
For security and privacy reasons,
he believes that the qualified requesters to attend this event should be people 
who are linked with at least two charities as he is, such as `UNICEF' and `Red Cross', 
that are involved in the humanity aid in Syria organized by the United Nations, 
i.e., `Unocha.Syria' in $\PG$,\footnote{\url{http://syria.unocha.org/}}.
The policy is defined as 
\[
 \begin{array}{l}
  \jmpu{\own}\up\markp{y_1} \ppath{\rt{donate}}\markp{y_5}(\nomp{Unocha.Syria}\wedge 
 \ppath {\rt{donate}^{\!-\!1}} \markp{y_3}\pu (\req\ \wedge\\ 
 \jmpu{\own}\up \markp{y_2} (\neg y_1 \wedge \ppath{\rt{donate}}(y_5 \wedge
 \ppath {\rt{donate}^{\!-\!1}}\markp{y_4}(\neg y_3\wedge \pu \req\ ))))) \\ 
 \end{array}
\]
The connections between the requester and Bob are shown in the right part of Figure~\ref{fig:cp}. 
Variables $y_1$ and $y_2$ mark two different charities;
so do $y_3$ and $y_4$ for the requester.
We notice that the charities that Bob is related to need not to be different from the ones of the requester.
Variable~$y_5$ guarantees that all these organizations have contributions to `Unocha.Syria'.

Since the public information and their relationships are extracted from external sources,
complicated relationship paths in $\PG$ as shown in this example
give rise to more meaningful and expressive access control policies.
%


\smallskip
\noindent\textbf{Scenario 3.}
In the third scenario in Section\ref{sec:intro}, Charlie only allows
his friends who work in the rival company of his employer to view his photo.
The policy is formally defined as below:
\[
 \jmpu{\own} (\upath{\rt{friend}} \req \wedge (\up \ppath{\rt{rival}}\pu \req)).
\]
Different from policies in the previous scenarios, this one requires that the owner and the
requester are linked through information in both $\UG$ and $\PG$.
More precisely, the sub-formula $\up\ppath{\rt{rival}}\pu$ regulates that the
qualified requester need to work for Company B's rival, i.e., Company A;
and the sub-formula $\upath{\rt{friend}}$ filters out the requester who is not a friend of Charlie.
We use a conjunction symbol to combine these two parts.
In Figure~\ref{fig:model}, only Alice is qualified as she is a friend of Charlie 
and she works for Company A.

\smallskip
\noindent\textbf{Scenario 4.}
In the fourth scenario, suppose that Bob wants to organize another fundraising event,
and he wants to invite people who used to participate in the same charities as him and their friends to attend the event.
The policy formula is specified as below:
\[
 \jmpu{\own}\up_{\prop{IsCharity}}\pu (\req\vee \upath{\rt{friend}}\req).
\]
In Figure~\ref{fig:model}, 
Alice is invited to participate this event since she is linked with Bob through a charity (UNICEF).
Moreover, Frank, Gabriele and Charlie can also receive the invitation due to their friendships with Alice.
Here, the path that links Frank (as well as Gabriele and Charlie) and Bob is composed
by both public information and users in the social network model.

\section{Using Category Relation in Access Control}
\label{sec:category}

In this section, we explore the category relation
among public information
and incorporate it in our hybrid logic
for the aim of concisely specifying access control policies
based on public information.   

\subsection{The Category Relation in Public Information Graph}
Let us first consider another scenario. 
In the model depicted in Figure~\ref{fig:model},
Charlie is linked with several kinds of sports including Basketball and Tennis. 
Alice is also a sport fan and her favorite one is Tennis, while Danny likes Volleyball.
Charlie has a photo depicting him playing tennis. 
He only wants his friends who are linked with Tennis to view it. 
The policy can be defined as
\[
\jmpu{\own}\upath{\rt{friend}}( \req \wedge (\up \nomp{Tennis})). 
\]
Since Alice likes Tennis, she can view the photo.
Now, Charlie decides to relax the restriction such that 
the qualified requester should be his friend who likes any kinds of sports.
He modifies his policy as follows:
\[
\jmpu{\own}\upath{\rt{friend}}( \req \wedge \up (\ppath{\rt{is\mbox{-}a}}\nomp{Sports})). 
\]
Relationship path $\ppath{\rt{is\mbox{-}a}}$ in the formula marks all the public
information that are in an $\rt{is\mbox{-}a}$ relation with Sports in $\PG$, e.g., Tennis.
However, this policy cannot achieve Charlie's goal. 
For example, Danny is not able to view this photo even he is supposed to be.
This is because Volleyball is not linked with Sports but Team Sports in $\rt{is\mbox{-}a}$ relationship 
as shown in Figure~\ref{fig:model}.
In order to grant access to Danny, Charlie again modifies the~policy as follows:
\[
\jmpu{\own}\upath{\rt{friend}}( \req \wedge \up (\ppath{\rt{is\mbox{-}a}}
\nomp{Sports}\vee \ppath{\rt{is\mbox{-}a}}\ppath{\rt{is\mbox{-}a}} \nomp{Sports})).
\]

\begin{figure*}[!t]
\centering
\includegraphics[scale=0.36]{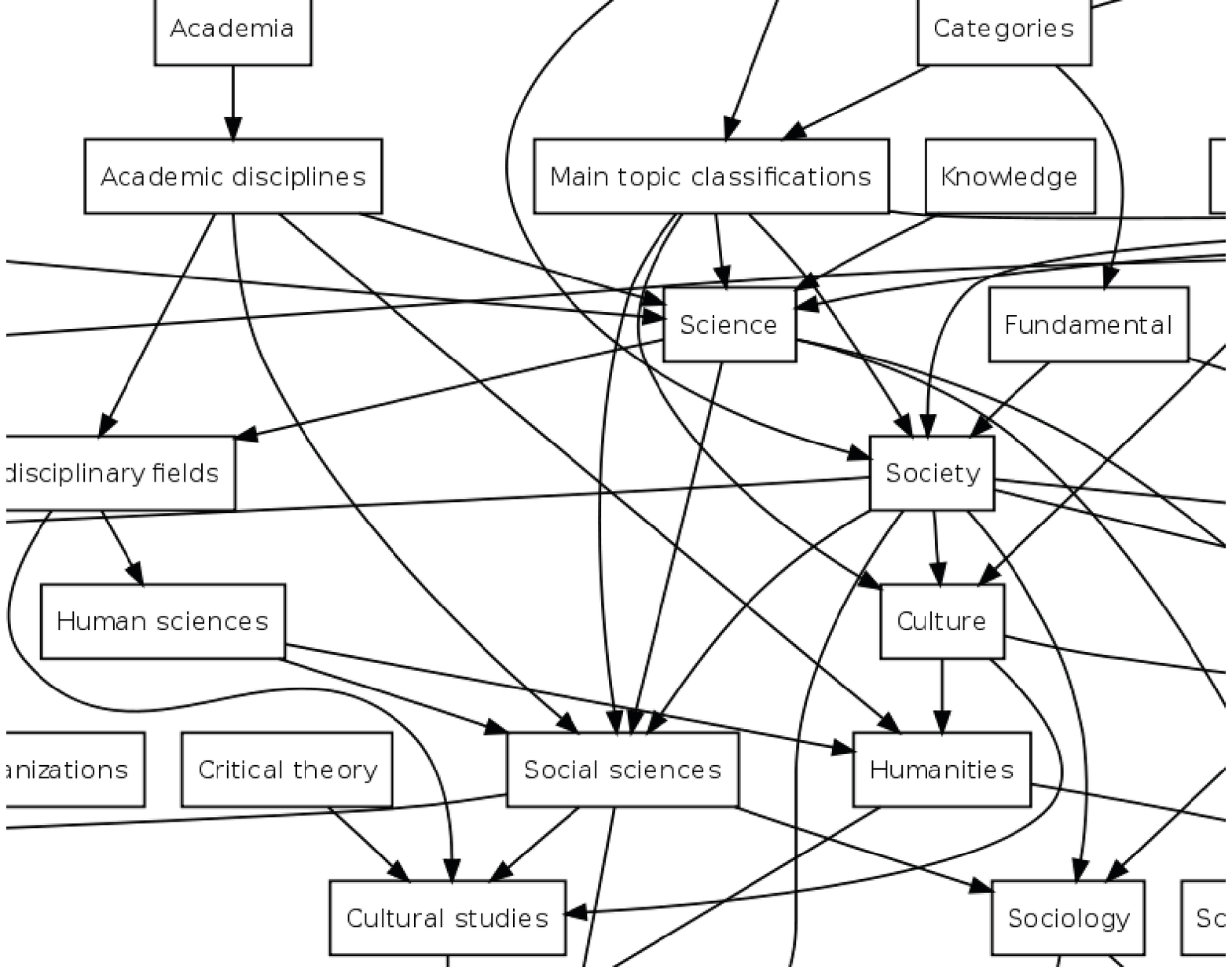}
\caption{Part of the category hierarchy of Wikipedia.\label{fig:category}}
\end{figure*}

However, there exists many public information related to Sports in the OSN and
defining a policy by enumerating all possible lengths is not an~acceptable solution.
In Wikipedia, articles are organized by means of categories
and all the categories form an acyclic graph.
Figure~\ref{fig:category} shows 
a part of the category graph of Wikipedia.\footnote{\url{http://en.wikipedia.org/wiki/Help:Categories}}
An article is under (at least) one category, some article can be the main article of a category.
For example, article basketball is under the category team sports, 
it is also the main article of the category basketball.
An article under a category is linked with the category's main article.
Actually, this is the $\rt{is\mbox{-}a}$ relationship among public information in $\PG$, 
we call it \emph{category relation}.
Since all categories of Wikipedia form an acyclic group (\emph{category graph}), 
public information together with $\rt{is\mbox{-}a}$ relationships among them compose an acyclic graph as well. 
For example, the subgraph in the dashed box in Figure~\ref{fig:model} is a tree.
Next, we integrate the category relation into our logic formula 
to express above policies in a concise way.


\subsection{Logic with the Category Relation}
In the model depicted in Figure~\ref{fig:model},
Charlie is linked with several kinds of sports including Basketball and Tennis. 
Alice is also a sport fan and her favorite one is Tennis, while Danny likes Volleyball.
Charlie has a photo that he wants to share with all his friends who like sports.
As depicted in the dash box of Figure~\ref{fig:model}, these kind of public information are organized by categories.
Instead of defining a policy to specify all the sports that are linked to users, 
we can directly use these category information to define policies.

To make use of the category relations among public information,
We first introduce a function on $\PG$ and a new symbol in our logic.
The~function $\it cf$ is formally defined as
\[
  \catf{\{\p{c}\}}= 
 \left \{ 
 \begin{array}{ll}
 \  {\{\p{c}\}} &\ \nexists\p{d} \mbox{ s.t. } (\p{d},\p{c})\in \rt{is\mbox{-}a}\\
 \bigcup\catf{\{\p{d}\}} & \ \forall \p{d} \mbox{ s.t. } (\p{d},\p{c})\in \rt{is\mbox{-}a} 
 \end{array} \right.
\]
The result of $\catf{\{\p{c}\}}$ contains $\p{c}$ and all its descendants in an acyclic graph 
based on $\rt{is\mbox{-}a}$ relationships in $\PG$.

In our hybrid logic, nominal $n$ can represent name of any public information in $\PG$. 
In order to refer to the node named $n$ as well as all its descendants in the formula,
we add a \emph{category nominal} $\catn{n}$ into our logic.
The syntax of formulas $\pf{}$ is extended as follows:
\[
\arraycolsep=2pt
\begin{array}{lll}
\pf{}&::=&\ t\mid \catn{n}\mid q\mid \neg \pf{}\mid \pf{1}\wedge \pf{2}\mid  
  \ppath{j}\pf{}\mid \jmpp{t}\pf{}\mid \markp{y}\pf{}\mid \pu\uf{}.
\end{array}
\]
%
The~semantics of $\catn{n}$ is
\[
\arraycolsep=2pt
\begin{array}{lll}
\model{\p{c}} \vDash \catn{n} &\mbox{iff}& \p{c} \in \catf{\VP(n)}\bigcup \VP(n).
\end{array}
\]

With the category nominal,
Charlie can easily redefine his policy in the previous example as 
\[
\jmpu{\own}\upath{\rt{friend}}(\req\wedge \up \catn{\nomp{Sports}}). 
\]
Now, all friends of Charlie who are related to any kind of sport
activities, such as Alice and Danny, can access the photo.

Similar to the ones with their contents from Wikipedia, public information from geography databases,
i.e., places, together with $\rt{is\mbox{-}in}$ relationships among them also naturally compose an acyclic graph.
Therefore, we are able to define policies to qualify the requester, 
such as ``only my friends who have ever been to Europe", 
in a concise way without listing different length of $\rt{is\mbox{-}in}$ relationship paths in $\PG$. 
Other types of hierarchical relationships on public information can also be investigated for the same purpose.

\section{Relationship Hierarchy}
\label{sec:hierarchy}
In this section, we extend our hybrid logic to
capture the hierarchy among different relationships,
enabling policy propagation in our access control scheme.


\subsection{Relationship Hierarchy}
Our social graph model supports multi-relationships.
As depicted in Figure~\ref{fig:model}, Gabriele and Danny are brothers and Alice and Danny are schoolmates.
In general, different relationships have different social strength.
Family-related relationships, such as spouse and parents, are normally considered stronger than professional relationships such as colleagues.
When an owner allows others who are in a certain relationship with him to view one of his resources,
those who are in a stronger relationships with the owner intuitively should be able to access the resource as well.
For example, if Alice allows her colleagues to view her education background, 
then her husband and parents should also be able to see it.

\begin{figure*}[!t]
\centering
\includegraphics[scale=0.6]{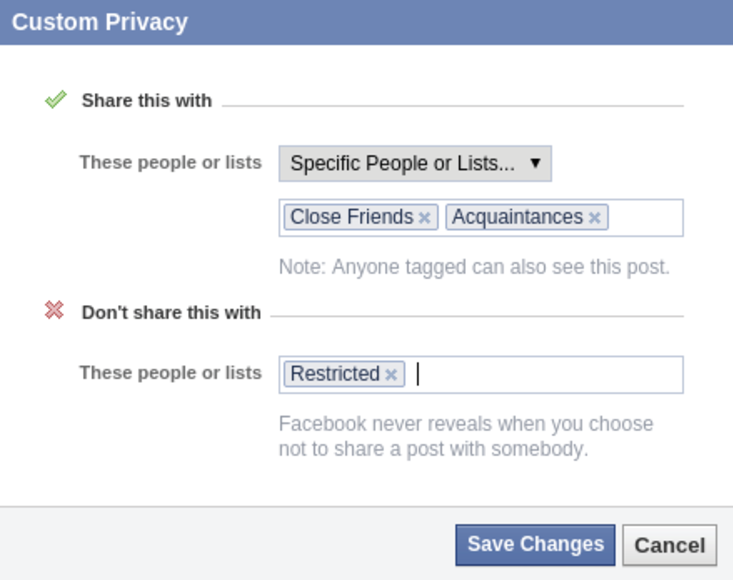}
\caption{Access control with close friends  acquaintances and restricted in Facebook.\label{fig:strengthaccess}}
\end{figure*}

In our hybrid logic, to express this kinds of policy, 
we can define a formula for each relationship type and connect these formulas together with the disjunction operator $\vee$.
The policy formula for the above example in our hybrid logic can be specified as
\[
 \jmpu{\own}\upath{{\it colleague}}\req \vee \jmpu{\own}\upath{{\it wifeof}}\req \vee \jmpu{\own}\upath{{\it childof}}\req.
\]
However, this solution is not ideal since it requires the owner to specify
the policy for all the intended relationships one by one.
It is very likely that the owner misses some relationships, thus the policy cannot fully capture his intention.
Therefore, we need a straightforward way to let the owner only specify one relationship in the policy 
and all the users who are in a stronger relationship with him can access the resource directly.
In fact, Facebook already allows a user to put his friends into three (smart) friend lists
including ``close friend'', ``acquaintances'' and ``restricted'' based on their social strength.
However, as depicted in Figure~\ref{fig:strengthaccess}, a Facebook user still needs to 
specify these lists in the audience selector (see Section~\ref{sec:motiv}) to control who can view his resource,
i.e., access control based on social strength is not implemented automatically in Facebook.

To express this kinds of policies in the hybrid logic, we first need to define a hierarchy on all the relationships
supported by the OSN.
This hierarchy can be built at a system level or a user level.
At a system level, OSN operators could regulate the order of relationship types with respect to their social strength.
On the other hand, different users may have different opinions about the strength of the relationships.
For example, some users believe that college friends are more important than colleagues from work
while some have the opposite opinion.
Therefore, OSNs could delegate this right to each user and let them freely define the relationship hierarchies themselves.

Here, for the sake of simplicity, we simply assume that the relationship hierarchy is defined at a system level.
This indicates that all users in the OSN will share the same relationship hierarchy.
The definition of the relationship hierarchy is given as follows.

\begin{defn}
A relationship hierarchy is defined as $(\UR, \leq)$,
where $\UR=\{\urt{1},\urt{2},\ldots,\urt{k}\}$ is the relationship type set and $\leq$ is a binary relationship on $\UR$
which is reflexive, antisymmetric and transitive.
\end{defn}

By its definition, a relationship hierarchy is a partially ordered set.
For two relationship types, $\urt{1}\leq \urt{2}$ indicates that $\urt{2}$ is a closer relationship
than $\urt{1}$.
Figure~\ref{fig:hierarchy} gives an example of the hierarchy.
In this example, spouse is considered the strongest relationship followed by close friends and family.
Note that the actual strength of the relationships is out of the scope of this work,
OSN operators can follow any theory from the area of sociology to construct the relationship hierarchy.


\subsection{Logic with Relationship Hierarchy}


To exploit the information in relationship hierarchy to specify our access control policies,
we introduce a new symbol $\rh{\upath{\urt{i}}}\uf{}$ into our syntax.
The syntax of the user formula is extended to:
\[
\arraycolsep=2pt
 \begin{array}{ccccccccccccccccccc}
    s &::= & m & \mid & x \\
  \uf{} &::= & s & \mid & p & \mid & \neg \uf{} & \mid & \uf{1}\wedge \uf{2} & \mid & \upath{\urt{i}}\uf{} &\mid&\rh{\upath{\urt{i}}}\uf{}& \mid & \jmpu{s}\uf{} &
  \mid & \marku{x}\uf{} & \mid & \up\pf{}.   \\
\end{array}
\]
The semantics of $\rh{\upath{\urt{i}}}\uf{}$ is defined below.
\[
\arraycolsep=2pt
\begin{array}{lll}
\model{\usr{a}} \vDash \rh{\upath{\urt{i}}}\uf{} &\mbox{iff}& \exists\ \usr{b}\in \U 
\mbox{ s.t. } (\usr{a},\usr{b})\in \urt{j} \mbox{ where } \urt{i}\leq\urt{j}\\
& &\wedge\ \model{\usr{b}} \vDash \uf{}\\
\end{array}
\]
Here, $\usr{b}$ can be in any relationship that is at least the same level of $\urt{i}$ with $\usr{a}$ defined in the relationship hierarchy.
To evaluate the policy, the relationship hierarchy should be included in the model $\MU$ as well.

\smallskip
\noindent\textbf{Example 1.}
Now, with the new operator, an owner could define a policy regulating that users who are at least his colleagues can view one of his resource as 
\[
 \jmpu{\own}\rh{\upath{\rt{colleague}}}\req.
\]
In addition, the hierarchy operator can be aligned together to express relationship path as well.
For example, the following policy means that the requester has to be 3-depth away from the owner and 
the relationship on each step has to be at least colleague:
\[
 \jmpu{\own}\rh{\upath{\rt{colleague}}}\rh{\upath{\rt{colleague}}}\rh{\upath{\rt{colleague}}}\req.
\]

\begin{figure*}[!t]
\centering
\includegraphics[scale=0.35]{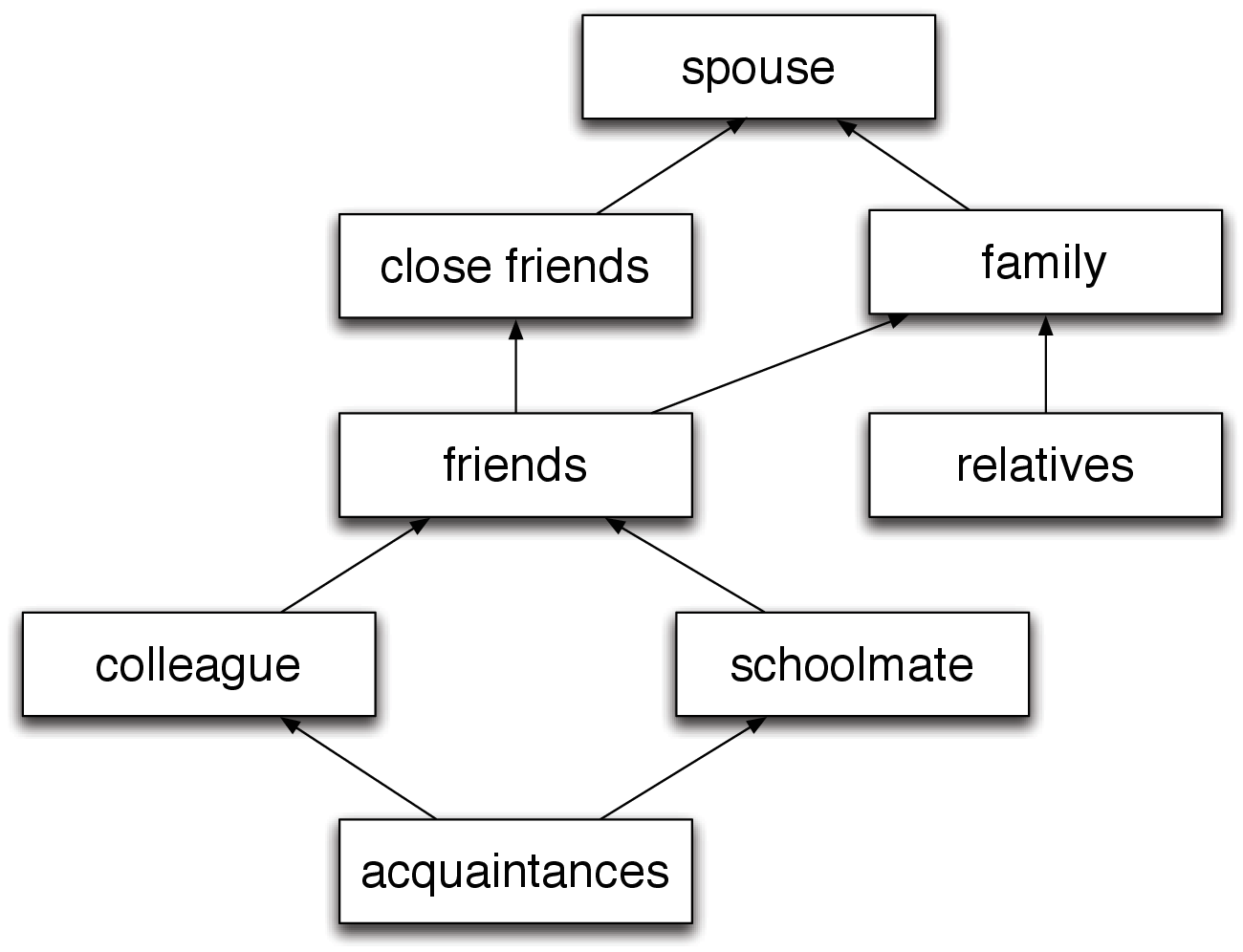}
\caption{A relationship hierarchy example.\label{fig:hierarchy}}
\end{figure*}

\smallskip
\noindent\textbf{Example 2.}
To give another example on how to use the hierarchical relationships, recall the social network depicted in Figure~\ref{fig:model},
suppose that Danny wants to share his interest, such as Volleyball, with his friends.
It is clear from Figure~\ref{fig:model} that only Charlie can view the information.
If Danny intends to share it with users who are also in stronger relationships with him, 
e.g., Eve (his wife) and Gabriele (his brother),
then the policy without using relationship hierarchy will be defined below,
where Danny has to explicitly enumerate all the relationships that he considers stronger than freind:
\[
 \jmpu{\own}\upath{{\it friend}}\req \vee \jmpu{\own}\upath{{\it husbandof}}\req \vee \jmpu{\own}\upath{{\it brotherof}}\req.
\]
Now, given the extended logic that supports hierarchical information,
Danny could simply redefine the policy in a more concise way:
\[
 \jmpu{\own}\rh{\upath{{\it friend}}}\req.
\]
Moreover, if Danny considers schoolmate a stronger relationship than friend 
which is different from the hierarchy presented in Figure~\ref{fig:hierarchy},
then Alice can access the resource as well.
In this case, instead of using the system level relationship hierarchy,
Danny could define his own relationship hierarchy 
with ${\it friend}\leq{\it schoolmate}$ specified.
In general, with the extension, our logic can support any hierarchical relationships when defining access control policies.

The main difference between relationship hierarchy and category relationship 
introduced in Section~\ref{sec:category} is the following:
relationship hierarchy is defined on relationships,
it can only grant access to users who are at the certain distance (specified in the policy) but in different relationships with the owner;
on the other hand, category relationship is defined on the nodes in public information graph
and it can represent paths of different length in a policy (through the recursively defined function~$\catf{\{\p{c}\}}$).
Further combination of the category relation and relationship hierarchy can be achieved as well,
which will give rise to a more powerful way to specify complicated policies in a simple form.

\smallskip
\noindent\textbf{Example 3.}
To give an example, in Figure~\ref{fig:model}, 
suppose that Bob wants to organize a fundraising event, 
he plans to invite all users who have involved in any fundraising event for Syria before,
and their friends (or stronger relationships than friends).
This policy exploits both category information related to public information graph
and relationship hierarchy related to social graph.
By combining the two extensions we have proposed,
Bob can simply define the policy as
\[
 \jmpu{\own}\up{\nomp{\catn{\nomp{Unocha.Syria}}}}\pu (\req\vee \rh{{\it friend}}\req).
\]
In Figure~\ref{fig:model}, Alice and Eve can see the invitation 
since they are directly involved in some fundraising events (through category relationship).
Besides, Frank, Gabirele and Charlie as friends of Alice can join as well.
Due to the power of relationship hierarchy, 
Danny can access the information since he is Eve's husband (following the relationship hierarchy in Figure~\ref{fig:hierarchy}).
On the other hand, without the extensions related to category and relationship hierarchy, 
to define a policy like this, Bob's policy formula will become much longer.
We conclude that both extensions improve the concision of our access control scheme.

\section{Information Reliability}
\label{sec:trust}

Owners define policies to control access to their resources.
However, in some cases, if the information in OSNs are not reliable,
malicious users can still gain access to some resources 
that they are not supposed to under certain policies.
For example, in Scenario~0 of Section~\ref{sec:example},
If an adversary is able to become friends with three friends of Eve, 
then he is able to gain the access.
Similarly in Scenario~3 of Section~\ref{sec:example},
a colleague of Charlie, who is also his friend, can maliciously specify that 
he works for the rival company in the OSN to access Charlie's sensitive photo.
As introduced in Section~\ref{sec:model}, 
our OSN model contains three parts, i.e., $\UG$, $\PG$ and two maps $\A$ and $\B$.
We discuss about their reliability one by one.

\smallskip
\noindent\textbf{Reliability of $\UG$.}
Information contained in $\UG$ are mainly users and their relationships.
Since a user can describe who he is in the OSN,
we only focus on users relationships.
To increase user relationships' reliability, we explore trust.
In contrast to the real life, 
trust between users in OSNs can be quantified, i.e., it has a value.
We first add trust values into $\UG$.
When $\usr{a}$ establishes an $\urt{i}$ relationship with $\usr{b}$, 
$\usr{a}$ will assign a trust value $\trva{ab}^{\urt{i}}$ to this relationship.
The edge from $\usr{a}$ to $\usr{b}$ is then defined as $(\usr{a},\usr{b},\urt{i},\trva{ab}^{\urt{i}})$.
Similarly, the edge from $\usr{b}$ to $\usr{a}$ is $(\usr{b},\usr{a},\urt{i}^{\!-\!1}, \trva{ba}^{\urt{i}^{\!-\!1}})$.
Note that $\trva{ab}^{\urt{i}}$ is only known to $\usr{a}$ and $\trva{ba}^{\urt{i}^{\!-\!1}}$ is only known to $\usr{b}$,
and these two values can be different.
We regulate that every trust value is in the interval $[0,1]$, the bigger the value is, more trust it represents.
We additionally introduce two new operators $\upath{\urt{i}}^{\rightarrow t}\uf{}$ and $\upath{\urt{i}}^{\leftarrow t}\uf{}$ 
into the user formula $\uf{}$ and their semantics are defined as follows.
\[
\begin{array}{lll}
\model{\usr{a}}\vDash \upath{\urt{i}}^{\rightarrow t}\uf{}  \mbox{~iff~} & \exists\ \usr{b} \in \U \ \mbox{s.t.~}  
                                       (\usr{a},\usr{b})\in \urt{i},
                                       \trva{ab}^{\urt{i}}\geq t\ \mbox{~and~} \\
 & \model{\usr{b}}\vDash \uf{}\\
 \model{\usr{a}}\vDash \upath{\urt{i}}^{\leftarrow t}\uf{} \mbox{~iff~} & \exists\ \usr{b} \in \U \ \mbox{s.t.~} 
		                       (\usr{b},\usr{a})\in \urt{i}^{\!-\!1},
		                       \trva{ba}^{\urt{i}^{\!-\!1}}\geq t\ \mbox{~and~} \\
& \model{\usr{b}}\vDash \uf{}                      
              
\end{array}
\]
When the requester is regulated to be linked with the owner through user relationships, 
trust can be put into the formula.
Now for the policy of Scenario~0, Eve can specify the formula as below: 
\[
 \jmpu{\own}\upath{\rt{friend}}_3^{\rightarrow 0.8}\req.
\]
To get an illegal access with the above formula, 
a malicious user needs to become friends with three users that Eve trusts ($t\geq0.8$).
Note that the way we integrate trust value into the user formula is simple.
There exist other methods, 
such as trust value can be evaluated on a whole relationship path.
How to extend our logic to support complicated trust requirements is part of our future work.

\smallskip
\noindent\textbf{Reliability of $\PG$.}
Different from users' information,
public information are imported from external databases and they are not operated by real users.
For example, Paris's information in Facebook is taken from Wikipedia
and the fact that it is in France can be extracted from public geography database.
Therefore, reliability of public information are guaranteed by these external sources
-- for instance, the reliability of Wikipedia pages and their connections can
be ensured by a community effort and users' reputation~\cite{AA07}.

\smallskip
\noindent\textbf{Reliability of $\A$ and $\B$.}
Some public information result in user relationships, 
for example, users who went to the same school are `schoolmates' or work in the same company are `colleagues'.
If the link between the qualified requester and this kind of public information 
are exploited by a policy,
then the owner who defines this policy 
can add the connection originated by the public information 
between the qualified requester and other users into the formula as well.
In this way, these other users can be treated as endorsing 
the connection between the requester and the public information.
In Scenario~3 of Section~\ref{sec:example}, 
besides working in the rival company, 
Charlie regulates that the qualified requester should have a certain number, e.g., 3, of colleagues 
who work in this rival company.
Moreover, he can also add trust to the formula.
The policy is defined as follows.
\[
\begin{array}{ll}
 \jmpu{\own}(\upath{\rt{friend}}^{\rightarrow 0.8}\req 
\wedge (\up\ppath{\rt{rival}}\markp{y}\pu(\req
\wedge\upath{\rt{colleague}}_{3}^{\leftarrow 0.7}\up y))).
\end{array}
\]
Now, in order to gain the access, the malicious user has to be trusted by Charlie ($t\geq0.8$) and
be colleagues with three other users who work in that company. 
Also, these three colleagues' trust value on the requester have to be at least 0.7.
Clearly, it is much harder for the adversary to succeed.

For policies exploiting public information that cannot result in user relationships,
endorsement (as well as trust) cannot be applied.
For example, in Scenario~1 of Section~\ref{sec:example}, 
the qualified requester needs to be linked to a location,
while in Scenario~2 Bob and the requester are connected through charities.
Similar to public information, the reliability of the links between some of these 
public information and users also depends on external services.
For example, in Facebook, a user is treated as having been to one location
if he used to publish a status or photo labeled with that location.
This location label is provided by ISP (Internet Service Provider) or GPS services.
A user's connection to a charity can be certified by the charity,
as the user normally gets tax benefit for his donations.
Again, we do not focus on the reliability of external services.

\section{Collaborative Access Control}
\label{sec:collaborative}

So far, we have assumed that the resource's access control policy can be only defined by its owner.
However, as introduced in Section~\ref{sec:logic}, 
a resource can be affiliated with several users, e.g., a photo tagged with several users,
and each of them should have the right to decide who can access the resource.
This is the so-called collaborative access control.
In this section, we aim to extend our model to support collaborative access control.

We first name all the users who are affiliated with a resource and are not the owner as the \emph{co-owners} of the resource.
We further use the set $O(r)$ to represent a resource $r$'s owner and co-owners.
If one co-owner of a resource wants to define a policy to allow only his friends of friends to view the resource,
then the policy formula is specified as $ \jmpu{\own}\upath{\rt{friend}}\upath{\rt{friend}}\req$.
For simplicity, we still use variable $\own$ in the formula to refer to one of the co-owners in $O(r)$.


With multiple policies on a resource, access control
conflicts can happen when deciding whether granting the access to a certain user or not.
Informally, a conflict means a user can access the resource under one policy but is forbidden by another.
For example, in the user graph depicted in Figure~\ref{fig:model}, 
suppose that Alice publishes a photo and tags her friends Bob and Gabriele in it.
Here, Alice is the owner while Bob and Gabriele are the co-owners of the photo.
We assume that Alice and Bob only allow their friends to view this photo 
and Gabriele wants users who are at least his friends to view it (see Section~\ref{sec:hierarchy}).
Their policy formulas as well as users who can access the photo, namely qualified users, are listed in Table~\ref{tab:accessusers}.
There are several access control conflicts.
For example, Eve can access the resource under Bob and Gabriele's policies 
but she is forbidden by Alice.
Note that the owner and co-owners of resource can always access the resource, 
and they are not included in the qualified users of each policy.

To formalize access control conflicts,
we first define the set of qualified users of a policy as the following.
\begin{defn}
 Given an access control policy $\phi$ that is defined by  a user $u$ on a resource $r$, i.e., $u\in O(r)$, 
 its set of qualified requesters is
 $\mathcal{QU}(\phi)=\{ u'\ \vert\ \MU,  \MP,  \A,  \B, u,  \g[\own\mapsto u, \req\mapsto u']\vDash\uf{}
 \wedge u'\notin O(r)\}$.
\end{defn}
Then, the \emph{conflict} on accessing a resource is defined as
\begin{defn}
 Given a resource with the set of access control policies defined on it, denoted by $\varPhi$.
 An access control conflict happens if there exists $u\in \mathcal{QU}(\phi)$ for a policy $\phi\in\varPhi$
 such that $u\notin\mathcal{QU}(\phi')$ for another policy $\phi'\in\varPhi$.
\end{defn}

Several works have been proposed to resolve conflicts
caused by collaborative access control (see Section~\ref{sec:related} for a short introduction), 
we can apply some of them within our scheme.
For instance, Hu et al.~\cite{HAJ13} proposed a few solutions for resolving
access control conflicts.
In their work, the so-called \emph{naive} solution is to only allow the common users
in the sets of qualified requesters to access the resource.
In the example of Table~\ref{tab:accessusers},
no one except for the co-owners can view the photo.
This shows that the \emph{naive} solution is too restrictive.
In addition, more sophisticated solutions based on voting schemes are proposed 
by Hu et al.~\cite{HAJ13} and others~\cite{SSP09}.
The voting scheme proposed in~\cite{HAJ13} contains two voting mechanisms, 
namely decision voting and sensitive voting.
For the decision voting, each co-owner is assigned a weight on his vote.
This weight can be equal for everyone or other rules may apply as well, such as the owner's vote has more weight than other co-owners'.
The final access control decision is made by accumulating all the owner and co-owners' votes.
If the final result is above a certain threshold, then the access is granted.
For the sensitivity voting, each user assigns a sensitivity level to the resource that he co-owns with others.
This means the scheme is resource based, 
i.e., a user can have a low sensitivity level on one resource but
a high sensitivity level on another resource.
Similarly, the final decision for the sensitivity voting is made by considering the total sensitivity level on the resource.
We notice that the decision voting and sensitivity voting can be combined together 
to further improve the process on resolving conflicts.

So far, we have considered conflicts at the requester level,
i.e., conflicts happen when different co-owners allow different users to access the resource.
In~\cite{SZPAM12}, Sun et al.\ considered conflicts at a policy level
and proposed an approach for resolving conflicts by combining trust relations in OSNs 
and preferential voting schemes. 
Under their consideration, a conflict happens when co-owners' policies are different.
In Figure~\ref{fig:model}, following the example in this section,
Alice, Bob and Gabriele co-own a photo.
Since the policies listed in Table~\ref{tab:accessusers} from them are different,
a policy-level conflict happens.
The solutions to resolve the requester-level conflicts can be naturally exploited to resolve the policy-level ones.
For instance, one naive solution would be: only the owner's policy is enforced on controlling the photo's access.
In this case, Gabirele's policy is ignored.

We notice that, in some cases, there are no policy-level conflicts but requester-level ones.
For example, in Table~\ref{tab:accessusers}, Alice and Bob have the identical policy,
thus there is no policy-level conflict between them.
On the other hand, as we discussed before, their policies still cause requester-level conflicts.
In some other cases, there may be no requester-level conflicts but policy-level ones.
For instance, suppose that in Figure~\ref{fig:model}
Alice and Charlie are tagged in a same photo when they watched a Tennis game at school several years ago.
Charlie wants to share this photo with his friends who like sports, i.e.,
$\jmpu{\own}\upath{{\it friend}}(\req \wedge \up\catn{\nomp{Sports}})$.
In Figure~\ref{fig:model}, except for Alice, only Danny is qualified.
Alice, however, only wants to share this photo with her schoolmates,
In Figure~\ref{fig:model}, only Danny can view the photo under Alice's policy.
There is no conflict at the requester-level since the only qualified requester is Danny.
However, Alice and Charlie's policies are obviously different which results in a policy-level conflict.
The relationship between these two types of conflicts deserves further investigations, 
we leave it as a future work.

\section{Comparison}
\label{sec:compare}

In this section, we compare our scheme with relationship-based access control schemes
in the literature~\cite{BFSH12,CFHKT09,CPS12} (see Table~\ref{tab:compare}).
%
%

The model of OSNs in~\cite{BFSH12} is the same as our user graph $\UG$,
but public information are not treated as entities in the model.
As a consequence, access control policies only make use of users' social representations.
On the other hand, it seems possible to express connections 
between users and public information through propositions in~\cite{BFSH12}.
For example, a proposition \emph{IsinParis} can be used to express the connection 
between a user and city Paris.
However, as mentioned in Section~\ref{sec:example}, 
each user will be affiliated with a large amount of attributes which is neither ideal or practical.
Moreover, policies that explore relationships between public information
(see examples in Section~\ref{sec:example}), cannot be captured by propositions.

The work proposed in~\cite{CFHKT09} does not explicitly take into account 
public information and their relationships.
However, this work has two interesting features.
First, in the OSN model, users' resources are treated as independent entities.
Relationships between users and resources are not restricted only to ownership,
e.g., the relationship between a user and a photo that he is tagged in 
is modeled as `photoOf' in their language.
Thus, collaborative access control is possible in their model.
Second, due to the fact that OSNs are modeled with semantic web technologies,
hierarchy information among users' relationships are naturally supported as well as
actions and resources, which make policy propagation possible.
For~example, if a user defines a policy to regulate the qualified requester to be his friends, 
then users who are in a closer relationship, such as `good friend', with him are also qualified.
In our work, we show how to perform policy propagation
based on a model of relationship hierarchy in our access control scheme (see in Section~\ref{sec:hierarchy}).
In addition, we used semantic relations among the public information in Section~\ref{sec:category}
to facilitate users to express their policies concisely.

Similarly, the scheme in~\cite{CPS12} does not take into account public information neither.
In this model, attributes of users are not represented.
Moreover, their policy language seems weaker than ours -- 
negation symbol only works with relationship paths, but not on nodes. 
Hence, policies such as ``all my friends but Alice can view my photo" cannot be expressed. 
On the other hand, this work has some its own features.
First, the OSN model treats resources as nodes which is similar to the one in~\cite{CFHKT09},
and actions that users performed on their resources are recognized as relationships.
For example, a user can regulate that only users who used to comment on a same photo 
as he did is able to poke him.
To support this in our access control model, 
we need to extend the social network model and treat users' resources as nodes as well.
Second, the authors propose a~simple solution through administrative policies 
for collaborative access control.
To achieve this in our model, we need to add a decision module in the model checking algorithm.

We also notice that the two schemes~\cite{CFHKT09,CPS12} can possibly treat public information as users' resources, 
i.e., modeled as nodes in their OSN model.
However, as we explained previously in Section~\ref{sec:model}, 
public information are extracted often from external databases, 
and relationships among them are different from the ones between users.
In our work, we apply the \emph{separation of concerns} principle
to model public information and their relationships separately from users and their social links.

\section{Discussion}
\label{sec:discuss}

We have shown that our scheme and its extensions can express fine-grained access control policies 
related to users and public information.
We have also shown how to deal with the problem of information reliability in OSNs
by incorporating endorsement and trust into our policy formulas.
There are still two other issues to discuss.

The first question is about the \emph{usability} of our scheme, especially for the non-experienced users --
whether a user can easily express a policy of his intention.
On one hand, relationship-based policies (e.g., friends, friends of friends) 
can be easily expressed in our scheme like the current access control schemes adopted by OSNs.
On the other hand, a group of qualified requesters under a sophisticated policy can be computed by OSNs, e.g.,
a Facebook user can directly get a list of his friends who have been worked in a company through Graph Search.
Besides, as shown in Figure~\ref{fig:smartlistaccess},
Facebook already implemented smart list for users to define fine-grained policies.
Therefore, we believe that our scheme can be supported as well.
Moreover, users can use visualization tools (e.g., see~\cite{AFYH09}) to learn whether their policies have been properly enforced.

The second is related to the \emph{availability} of user information in OSNs.
As privacy raises serious concerns in OSNs, users might not be willing to share too much information.
As a consequence, some eligible users can be filtered out by a policy due to the lack of their information in the OSN.
However, the main purpose of OSNs is for people to express themselves and socialize with other users --
more information a user shares, more benefits he will gain from the OSN.
On the contrary, a user keeps more privacy if he shares less information.
There is always a balance between information sharing (or utility) and privacy.
What we focus in this paper is to explore the information shared by users in OSNs to 
express fine-grained access control policies.
Thus, we consider availability of user information in OSNs orthogonal to our proposal.

\section{Conclusion and Future Work}
\label{sec:conclusion}

In this paper, we have first identified a new type of access control policies 
that are meaningful but have never been addressed in the literature.
Namely, users in OSNs can express access control requirements
not only based on their social relations but also on their connections through public information.
Then we defined an OSN model containing users and public information,
based on which we proposed a hybrid logic to define access control policies. 
We gave a number of policies based on public information and 
formulated them formally and precisely in our proposed logic.
We further used category relations among public information and relationship hierarchy
to extend our logic and make it more practical.
In addition, we also showed how to extend our model and logic to deal with
unreliable information and collaborative access control in OSNs. 

In the future, besides studying the relationship between requester-level and policy-level conflicts 
related to collaborative access control framework (see Section~\ref{sec:collaborative}),
we plan to improve the expressiveness of our model
by integrating user resources~\cite{CFHKT09,CPS12}.
As resources are different from users, modeling resources explicitly
may address more expressive policies.
The connections between resources and public information will be interesting to study as well.
Secondly, we plan to develop a Facebook App to support our access control models.
This App should guide users to use public information within Facebook
to express their intentions on control their resources' access.
The main feature of this App is to give users a way to organize their friends into different lists or groups
by exploring different public information.
Besides, a visualization tool, similar to the one of Anwar and Fong~\cite{AF12}, will be developed 
as part of the App to help users to find and evaluate who else in Facebook can access his resources
under his policies.

\bibliographystyle{elsarticle-num}
\bibliography{pig_journal}

\newpage
\begin{table*}
\centering
\scalebox{1}{
\begin{tabular}{|c|c|c|}
  \hline
  \mbox{(co-)owner} & Policy formula $\phi$ &  Qualified users\\
  \hline
  \hline
  Alice & $\jmpu{\own}\upath{\rt{friend}}\req$ & Frank, Charlie\\
  Bob & $\jmpu{\own}\upath{\rt{friend}}\req$ & Eve\\
  Gabriele & $\jmpu{\own}\rh{\upath{\rt{friend}}}\req$ & Eve, Danny\\
  \hline
  \end{tabular}
  }
  \caption{(Co-)owners with their policies and users who can access the resource.}
  \label{tab:accessusers}
\end{table*}

\begin{table*}
\centering
\scalebox{1}{
\begin{tabular}{|c||c|c|c|c|}
  \hline
            &\cite{BFSH12} & \cite{CFHKT09} & \cite{CPS12} & This paper \\
  \hline
  \hline
  Multi-relationship type       &  \checkmark        & \checkmark               & \checkmark         & \checkmark  \\
  \hline
  User attributes               &  \checkmark        & \checkmark               &                    & \checkmark  \\
  \hline
  Public information            &                    &                          &                    & \checkmark  \\
  \hline
  Trust                         &                    & \checkmark               &                    & \checkmark  \\
  \hline
  User-resource relation        &                    &                          &  \checkmark        &             \\
  \hline\hline
  Relationship depth            &  \checkmark        & \checkmark               & \checkmark         & \checkmark  \\
  \hline
  Topology-based policy         &  \checkmark        &                          &                    & \checkmark  \\
  \hline   
  Policy propagation            &                    & \checkmark               &                    & \checkmark  \\
  \hline                                
  \end{tabular}
  }
 \caption{Comparison of access control schemes for OSNs.}
\label{tab:compare}
\end{table*}

\newpage

\end{document}